\documentclass[aps,showpacs]{revtex4}
\usepackage{epsfig}

\begin{document}
\title{Remarks on the shape transition from spherical to deformed gamma unstable nuclei}

\author{A. A. Raduta$^{a),b)}$, A. C. Gheorghe$^{(b)}$ and Amand Faessler$^{c)}$}

\address{
$^{b)}$Department of Theoretical Physics and Mathematics,Bucharest University, POBox MG11,
Romania}
\address{$^{a)}$Institute of Physics and Nuclear Engineering, Bucharest, POBox MG6, Romania}
\address{$^{c)}$Institut fuer Theoretische Physik, der Universitaet Tuebingen,
auf der Morgenstelle 14, Germany}

\begin{abstract}
Energies and transition probabilities for low lying states in $^{134}$Ba and $^{104}$Ru were calculated within
a hybrid model. The ground and the first $2^+$ states are described
alternatively as a harmonic and
anharmonic vibrator states while the remaining states as states with
 $E(5)$ symmetry. Results for $^{134}$Ba are compared with those predicted  by
 some other methods. One concludes that a gradual setting of the 'critical'
 potential yields a better agreement with the experimental data.
 Very good agreement with the data is obtained for
 $^{104}$Ru. Comparing the present results with those of E(5) symmetry,
 it is conspicuous
 that the present formalism add corrections to the E(5) formalism
 by bringing the predictions closer to the experimental data.
 Analytical relationship between the states with $U(5)$ symmetry and those
 given by the $E(5)$ description is established.
\end{abstract}

\maketitle

\section{Introduction}
\label{sec:level1}
Since the liquid drop model was developed \cite{Bohr}, the
quadrupole shape coordinates  were widely used  by both
phenomenological and microscopic formalisms to describe the basic properties of
nuclear systems. Based on these coordinates, one defines quadrupole
boson operators in terms of which model Hamiltonians and transition operators
are defined. Since the original spherical harmonic liquid drop model was able
to describe only a small amount of data for spherical nuclei, several
improvements have been added. Thus, the Bohr-Mottelson model was generalized by
Faessler and Greiner\cite{GrFa}
in order to describe the small oscillations around a deformed shape which
results in obtaining a flexible model, called vibration rotation model,
suitable for the description of deformed nuclei. Later on \cite{Gneus} this picture was
extended by including anharmonicities as low order invariant  polynomials in the
quadrupole coordinates. With a suitable choice of the parameters involved in the model
Hamiltonian the equipotential
energy surface may exhibit several types of minima \cite{Hess} like spherical,
deformed prolate, deformed oblate, deformed triaxial, etc.
 To each equilibrium shape, specific properties for excitation energies and electromagnetic transition
 probabilities show up. Due to this reason,
one customarily says that static values of intrinsic coordinates determine a phase for the
nuclear system. A weak point of the boson description with a complex anharmonic Hamiltonian consists of the large number of the structure parameters which are
to be fitted.  A much smaller number of parameters is used by the
coherent state model (CSM) \cite{Rad1} which uses a restricted collective space generated through angular momentum projection by three deformed orthogonal functions of coherent type. The model is able to describe in a realistic fashion transitional and well deformed nuclei of various shapes including states of high and very high angular momentum. Various extensions to include other degrees of freedom like isospin
\cite{Rad2}, single particle\cite{Rad3} or octupole degrees\cite{Rad4} of freedom have been formulated\cite{Rad5}.
  
It has been noticed that a given nuclear phase may be
associated to a certain symmetry. Hence, its properties may be described with
the help of the irreducible representation of the respective symmetry group.
Thus, the gamma unstable nuclei can be described by the $O(6)$ symmetry
\cite{Jean}, the gamma triaxial nuclei by the rigid triaxial rotor $D2$ symmetry
\cite{Filip},
the symmetric rotor by the $SU(3)$ symmetry and the spherical vibrator by the
$U(5)$ symmetry.
Thus, even in the 50's, the   symmetry properties have been greatly appreciated.
 However, a big push forward was brought by the interacting boson
 approximation
(IBA) \cite{Iache,Iache1}, which succeeded to describe the basic properties of a large number of
nuclei in terms of the symmetries associated to the system of quadrupole (d) and
monopole (s) bosons  which  generate a $U(6)$ algebra. The three
limiting symmetries $U(5)$, $O(6)$, $SU(3)$ mentioned above, are dynamic symmetries
for $U(6)$. Moreover, for each of these symmetries a specific group reduction
chain  provides the quantum numbers characterizing the states, which are suitable
  for a certain region of nuclei. Besides  the virtue of unifying the group
  theoretical descriptions of nuclei exhibiting different symmetries,
  the procedure defines very simple reference pictures for the limiting cases. For nuclei
 lying close to the region characterized by a certain symmetry,
 the perturbative corrections are to be included.

A nice classification scheme was provided by Casten \cite{Cast}, who placed all
nuclei on the border of a symmetry triangle. The vertices of this triangle symbolize the
$U(5)$ (vibrator), $O(6)$ (gamma soft) and $SU(3)$ (symmetric rotor), while the legs
of the triangle denote the transitional region. Properties of nuclei lying far
from vertices are difficult to be explained since the
states have some characteristics of one vertex while some others are easy to be
described by using the adjacent symmetry. When the mixture of the extreme
symmetries is maximal, one says that the critical point for the transition from
one phase to another has been reached. In Ref. \cite{Gino,Diep}, it has been proved
that on the $U(5)-O(6)$ transition leg there exists a critical point
for a second order phase transition while the
$U(5)-SU(3)$ leg has a first order phase transition. 

Recently, Iachello \cite{Iache2,Iache9} pointed out that these critical points
correspond to
distinct symmetries, namely $E(5)$ and $X(5)$, respectively. For the critical value
of an
ordering parameter, energies are given by the zeros of a Bessel function of half integer and irrational indices, respectively.

It is worth mentioning that this procedure, is frequently used for quantizing
the classical electromagnetic field by confining it inside a sphere of radius 
$R$
on whose surface the solution of the field equation vanishes. Moreover, the magnitude of
the radius determine the photon energy.

Remarkable is the fact that the experimentalists  found representatives for
the two
symmetries. To give an example, the relevant data for $^{134}$Ba \cite{Zam}
and $^{152}$Sm \cite{Zam1}
suggest that they are close to the $X(5)$ and $E(5)$ symmetries, respectively.
Another candidate for E(5) symmetry, proposed by Zamfir {\it et al.} \cite{Zam2}
is $^{102}$Pd. Using a simple IBA Hamiltonian, in Ref.\cite{Da-li}, the low lying
spectrum of
$^{108}$Pd is realistically described. Comparing the E(5) predictions with the
experimental data concerning energy ratios in the ground band and the normalized
E2 transition probabilities for the states $4^+$ and $0^+_2$,
one concludes that this nucleus is a good E(5) candidate.
However, in order to decide which  Pd isotope is closer to an E(5) behavior, further
investigations are necessary.
A systematic search for E(5) behavior in nuclei has been reported in Ref.\cite{Clar}.

Short after the pioneering papers concerning critical point  symmetries
appeared, some other attempts have
been performed, using other potentials like Coulomb, Kratzer \cite{Fort} and
Davidson potentials \cite{Bona}. These potentials yield also Schr\"{o}dinger
solvable equations and the corresponding results may be interpreted in terms of
symmetry groups.

In this paper we refer to the $E(5)$ symmetry and try to get a consistent
description of some experimental data for $^{134}$Ba and $^{104}$Ru.
The basic idea of our work consists of the fact that going from the spherical
vibrator ($U(5)$) to the $O(6)$ like nuclei, by increasing the 'order parameter',
energy levels do not reach the critical behavior
at the same time i.e., for the same value of the above mentioned parameter.

The fact that the critical point might be state dependent has
been considered by several authors in different contexts. In what follows we shall
mention few relevant studies.
~a)~ The dependence of the critical points on angular momentum and how this dependence
influences the energies has been analyzed in Ref. \cite{Bona} for the
Davidson \cite{Davi} potential. ~b)~ In Ref.\cite{Rad86},
we analyzed the A dependence of the spectroscopic properties of Gd isotopes
within the CSM approach. The conclusion  was that the transition from spherical
(SU(5)) to deformed (SU(3)) phases takes place gradually, i.e. not simultaneously
for all energy levels. This feature has been analyzed in great details using both
the CSM predictions
and the experimental data. Therein, we stated that the critical behavior concerning the above mentioned
phase transition is met by $^{154}$Gd.
~c)~ A similar conclusion concerning the
dependence of the phase transition on the nuclear system state was reached in
Ref.\cite{Rad98}, where some isotopes of Xe and Ba have been described within
the CSM and
the triaxial rotation-vibration model (TRVM). Indeed, the ground state
$0^+_g$ predicted by CSM behaves like a gamma unstable state while the state $10^+_g$
has a maximum gamma probability for $\gamma=45^0$. In gamma band, the model
state  for
$2^+_{\gamma}$ describes a triaxial nucleus ($\gamma =30^0$) while increasing
the spin,
the probability distribution in the variable $\gamma$ goes gradually to the situation when it has two
equal maxima at $0^0$ and $60^0$. The $\gamma$ asymmetric features are
described by mixing the
ground and $\gamma$ model states of equal angular momenta through the
anharmonicities involved in the boson Hamiltonian.
~d)~ The triaxial rotor Hamiltonian, with the moment of inertia depending on
the nuclear deformation  $\beta$ in the manner predicted by the liquid drop
model \cite{Tur}, has
eigenvalues which are functions of $I$ (angular momentum) and $\beta$. For $I<8$,
the energy has a minimum in $\beta=0$, while
for $I\geq 8$ it has a deformed minimum. Therefore a nuclear system described
by such a Hamiltonian
stays in a spherical phase if $I<8$, and in a deformed phase for $I\geq 8$.
~e)~ A much more complex situation, when a triaxial rotator is cranked around an arbitrary
axis is studied in Ref.\cite{Gheor}. It is shown that the phase portrait depends strongly on the constraint,
i.e. on angular momentum of the state.
~f)~ In Refs.\cite{Jolie1,Jolie2} it was
pointed out that in nuclei exhibiting U(5) and O(6) shape coexistence, due to
the intruder
states, the components having the two symmetries respectively,
are selectively mixed, which results in having,
for the given nucleus, states with  pure symmetries and states where the two
symmetries are
mixed with each other.
Very strong arguments supporting the state dependence of the critical point is
also brought by
microscopic descriptions.
~g)~ It is well known that a nuclear rotating system may
exhibit a "superconducting" phase
up to a certain critical angular momentum when a nucleon pair is broken.
From this angular momentum on,
one develops a new band of a two quasiparticle nature, having properties which differ
essentially from those
characterizing the states of lower angular momenta\cite{Plo}.
~h)~ According to Ref.\cite{Garro} the phase transition from a reflection symmetry
to a static octupole deformed phase is taking place in an excited state.
~i)~ Moreover, in Ref.\cite{Raio} several cases were depicted, where   nuclear
systems may
stay in a phase with a good reflection symmetry for any angular momentum in
the ground band, but at a certain angular momentum in an
 excited band, the static octupole deformation is set on.
All examples mentioned above plead coherently for the idea adopted here saying
that
nuclear systems may change the phase when is promoted in an excited state.

In order to see the effect of angular momentum dependence of the critical
points on both energies and $E2$ transitions, we suppose that the ground state
( $0^+$) and the first excited  $2^+$ state are described by a spherical
vibrator, while the remaining states correspond to a different potential energy,
mocked up through a five dimensional infinite square well.
With this assumption we calculate the excitation energies in units of the
first $2^+$ state excitation energy as well as the transition probabilities
of the low lying states.
At  the second step, we introduce some anharmonicities in the structure of
the states $0^+$ and $2^+$.
A certain mathematical relationship between the states of the $U(5)$ and $E(5)$
symmetries are pointed out.

The aim sketched above is accomplished according to the following plan.
In Section 2 the results for U(5) and E(5) symmetries are shortly presented.
Energies and transition probabilities will be calculated in Section 3 by
supposing that the first two states are given by a spherical vibrator while
the remaining ones by the solutions of the Schroedinger equation for a
five-dimensional infinite square well.
Also, results for the situation when the ground state and the first excited state involve
some anharmonicities, are given. 
In Section 4, the connection between the spherical vibrator states and the E(5)
basis states is established. The final conclusions are drawn in Section 5.
\section{Brief description of the spherical vibrator and E(5) symmetries}
Written in the intrinsic frame of
reference, the original Bohr-Mottelson Hamiltonian has the expression:
\begin{equation}
H=-\frac{\hbar^2}{2B}\left[\frac{1}{\beta^4}\frac{\partial }{\partial \beta}
\beta^4 \frac{\partial }{\partial \beta}+\frac{1}{\beta^2\sin {3\gamma} }
\frac{\partial}{\partial \gamma}\sin{3\gamma}\frac{\partial}{\partial \gamma}
-\frac{1}{4\beta^2}\sum_{k=1,2,3}\frac{Q_k^2}{\sin^2(\gamma-\frac{2}{3}\pi k)}
\right]+V(\beta,\gamma),
\label{Has}
\end{equation}
where the dynamic deformation variables are denoted by $\beta$ and $\gamma$
while the intrinsic angular momentum components by $Q_i$, with $i=1,2,3$.
If the potential energy term is depending only on the beta variable
\begin{equation}
V(\beta,\gamma)=U(\beta),
\end{equation}
the eigenvalue equation associated to $H$ (\ref{Has}) can be separated in 
two parts, one
equation describing the beta variable and the other one the gamma
deformation and the Euler angles
$\Omega=(\theta_1,\theta_2,\theta_3)$. The equation in $\beta$ is:
\begin{equation}
\left[-\frac{1}{\beta^2}\frac{\partial }{\partial \beta}\beta^4
\frac{\partial }{\partial \beta}+\frac{\Lambda}{\beta^2}+u(\beta)\right ]f(\beta)
=\epsilon f(\beta),
\label{epsf}
\end{equation}
where   $ \Lambda$ is the eigenvalue of the Casimir operator of the $SO(5)$ group.
This is related with the seniority quantum number $\tau$,  by $\Lambda=\tau(\tau+3)$.
The 'reduced' potential $u(\beta)$  and energy $\epsilon$ are defined as:
\begin{equation}
E = \frac{\hbar ^2}{2B}\epsilon,\;\;\;U= \frac{\hbar ^2}{2B}u.
\label{EU}
\end{equation}
where $E$ denotes the eigenvalue of the Hamiltonian $H$ corresponding to the potential $U(\beta)$.
A full description of the eigenstates of the Bohr-Mottelson Hamiltonian
satisfying the symmetry $U(5)\supset SO(5)\supset SO(3)\supset SO(2)$,
may be found in Refs.\cite{Gheor}. In particular, the solution of the radial equation
(\ref{epsf}) with $u(\beta)=\beta^2$ is easily obtained by bringing first  Eq.(
\ref{epsf}) to the
standard Schr\"{o}dinger form by changing the function f to $\psi$ by:
\begin{equation}
\psi(\beta)=\beta^2 f(\beta).
\label{psi}
\end{equation}
The equation obeyed by the new function $\psi$, is:
\begin{equation}
\frac{d^2{\psi}}{d\beta^2}+
\left[\epsilon-\beta^2-
\frac{(\tau+1)(\tau+2)}{\beta^2}\right]\psi=0.
\label{Sch}
\end{equation}
This equation is analytically solvable. The solution is:
\begin{eqnarray}
\psi_n(\beta)&=&\sqrt{\frac{2(n!)}{\Gamma(n+\tau+5/2)}}
L^{\tau+3/2}_n(\beta^2)\beta^{\tau+2}\exp(-\beta^2/2), \\
\epsilon_n&=&2n+\tau+5/2,\;\;\;n=0,\;1,\;2,\;...;\tau=0,1,2,3,...
\label{spectr}
\end{eqnarray}
where $L_n^{\nu}$ denotes the generalized Laguerre polynomials. The number of
polynomial nodes
is denoted by $n$ and is related to the number of the quadrupole bosons ($N$) in
the state, by: $N=2n+\tau$.
Consequently, the initial equation (\ref{epsf}) has the solution
\begin{equation}
F_{n\tau}=\beta^{-2}\psi_{n\tau}\;.
\end{equation}
The spectrum, given by Eq. (\ref{spectr}), corresponds to the unitary representation of
the $SU(1,1)$ group having
the Bargman index $k=(\tau+5/2)/2.$ The standard generator for $SU(1,1)$ are:
\begin{equation}
K_0=\frac{1}{4}H_0,\;\;\;\;K_{\pm}=\frac{1}{4}\left[\frac{(\tau+1)(\tau+2)}{\beta^2}-(\beta\pm\frac{d}{d\beta})^2\right],
\label{su11}
\end{equation}
where
\begin{eqnarray}
H_0&=&-\frac{d^2}{d\beta^2}+\beta^2 +\frac{(\tau+1)(\tau+2)}{\beta^2},
\nonumber\\
H_0\psi_n &=& \epsilon_n\psi_n.
\label{H0}
\end{eqnarray}
The commutation relations of $H_0$ are:
\begin{equation}
\left[K_-,K_+\right] = -\frac{1}{2}H_0,\;\;
\left[K_{\pm},H_0\right] = \pm 4K_{\pm}.
\label{comut}
\end{equation}

If $u(\beta )=\beta ^{2}+g\beta ^{-2}$, then (2.6)-(2.11) hold for $\tau $
replaced \cite{Bona} by 
\begin{equation}
\tau ^{\prime }=-\frac{3}{2}+\left[ \left( \tau +\frac{3}{2}\right)
^{2}+g\right] ^{1/2}.  \label{dav0}
\end{equation}

Now, let us turn our attention to the situation considered by Iachello in
Ref.\cite{Iache2}, where
the potential term associated to the spherical to gamma unstable shape transition
 is so flat that it can be mocked up as a infinity square well
\begin{eqnarray}
u(\beta)=\left\{\matrix{0,&\beta\leq \beta_w,\cr
                       \infty, & \beta > \beta_w.}\right .
\label{udeb}
\end{eqnarray}
A more convenient form for the equation in $\beta$, is obtained through the function
transformation:
\begin{equation}
\varphi(\beta)=\beta^{3/2}f(\beta),
\label{fi}
\end{equation}
The equation for $\varphi$ is
\begin{equation}
\frac{d^2\varphi}{d\beta^2}+\frac{1}{\beta}\frac{d\varphi}{d\beta}+
\left[\epsilon-u(\beta)-\frac{(\tau +3/2)^2}{\beta ^2}\right]\varphi=0.
\label{diffi}
\end{equation}
Changing the variable $\beta$ to $z$ by
\begin{equation}
z=k\beta ,\;\;\; k=\sqrt{\epsilon}
\label{zsik}
\end{equation}
and denoting with $\widetilde {\varphi}(z)=\varphi(\beta)$  the function of the new
variable, one arrives at:
\begin{equation}
\frac{d^2\widetilde{\varphi}}{dz^2}+\frac{1}{z}\frac{d\widetilde{\varphi}}{dz}
+\left[1-
\frac{(\tau+3/2)^2}{z^2}\right]\widetilde{\varphi}=0.
\label{fitild}
\end{equation}
This equation is analytically solvable, the solutions being the Bessel
functions of half
integer order, $J_{\tau+3/2}(z)$. Since for $\beta>\beta_w$ the function
$\widetilde{\varphi}$ is equal to zero, the continuity condition requires that
the solution inside the well must vanish for the value of $\beta$ equal to $\beta_w$.
This, in fact, yields a quantized form for the eigenvalue $E$. Indeed,
let $x_{\xi,\tau}$ be the zeros of the Bessel function $J_{\nu}$ :
\begin{equation}
J_{\tau +3/2}(x_{\xi,\tau})=0,\;\;\xi=1,2,...;\tau=0,1,2,...
\label{zero}
\end{equation}
Then, due to the substitution introduced in Eqs.(2.12) and (2.4) one obtains:
\begin{equation}
E_{\xi,\tau}=\frac{\hbar^2}{2B}k^2_{\xi,\tau},\;\;k_{\xi,\tau}=\frac{x_{\xi,\tau}}{\beta_w}.
\label{Exi}
\end{equation}
Concluding, the differential equation for the beta deformation corresponding to an infinite well
potential provides the energy spectrum given by Eq.(\ref{Exi}) and the wave functions:
\begin{equation}
f_{\xi,\tau}=C_{\xi,\tau}\beta^{-3/2}J_{\tau+3/2}(\frac{x_{\xi,\tau}}{\beta_w}\beta),
\label{fxi}
\end{equation}
where $C_{\xi,\tau}$ is a normalization factor.

\section{ A hybrid model for the U(5) to O(6) transitional nuclei.}
\label{sec:level3}
Customarily, the extreme values for the energy function associated to a
model Hamiltonian
to be used for the description of the phase transition between two distinct
symmetries regime is determined variationally for each state of the spectrum.
As examples we mention the variational moment of inertia (VMI) model
\cite{Mari},
the Lipas-Holmberg equation \cite{Hol} for the ground band energies in transitional nuclei,
the solutions of liquid drop equation with Davidson potential \cite{Bona},
cranking formalism for a many body Hamiltonian \cite{Rad11}.
In Ref.\cite{Iache2}, it is suggested that a five dimensional well potential,
corresponding to the E(5) symmetry,
may simulate a critical flat potential in the beta variable. Indeed, one can
imagine a family of potentials, in the beta variable, depending on an order
parameter $\eta$ having the following property. Potentials with the order parameter smaller than the critical value  ($\eta_c$)
have spherical minimum while those for $\eta > \eta_c$ exhibit a deformed minimum.
It is remarkable the fact that the states with E(5) symmetry provide energy ratios and
normalized B(E2) values, for a transition operator linear in $\beta$, which do
not depend on the width of the potential well. Note that in Ref.{Iache2}, the
critical potential corresponding to $\eta_c$ is simply mocked by the five dimensional
well, but ignoring the process of how the order parameter is changed so to reach the value
$\eta_c$. The change of $\eta$ toward $\eta_c$ might be caused by modifying the
number of neutrons or by promoting a given nucleus in a state of higher angular momentum, or by
changing both the neutron number and angular momentum. As a matter of fact the lettest
version was considered in Ref.\cite{Rad86}. Indeed, therein we found out that
for a given angular momentunm the critical behaviour is reached in a certain
isotope while for a higher angular momentum the critical features appear in the next isotope.

Restrictions required for energy extreme points to exist define in the Hamiltonian
parameter space, regions characterized by the same phase space portrait for the
extreme points. That means that the phase space coordinates minimal values have
similar dependence on the structure coefficients for all points of parameter space
in the selected region. Due to this feature the subset of the parameter space with
this property is called "phase" of the nuclear system. Two distinct phases are
separated by curves reached by changing the structure parameters. The points of separatrices
are called critical points. When two curves bordering the same phase
intersect each other the common point defines the so called bifurcation point.
Depending on the initial conditions
the system may evolve toward one of more than two possible phases.
A complete analysis of the nuclear phases associated to some particular
phenomenological Hamiltonians have been presented in Refs. \cite{Casta,Rad12,Aria1}.

It is noteworthy the fact that in order to describe several nuclear phases
with a single Hamiltonian one needs to include anharmonicities\cite{Rad20}.
Among them the third order boson Hamiltonian plays a special role. Indeed,
is this term the one which causes instabilities for the equation of motion for the intrinsic degrees of freedom, $\beta$ and $\gamma$ \cite{Bar1,Bar2}.
In the model used by Iachello \cite{Iache2}, there is no polynomial
anharmonicity involved. By contrary, from the Bohr-Mottelson Hamiltonian for a
harmonic liquid drop, one drops out the harmonic potential energy, which is
the quadratic invariant $\beta^2$. As a result, one remains only with the
kinetic energy operator whose eigenstates can be factorized, the radial factor
being proportional to a Bessel function of half integer order. By choosing an
infinite wall for the potential in beta, the boundary condition provides
a quantized expression for energy. 
Note that such a system is 
quite soft against any deformation process indeed, since there is no surface 
tension term and, on the other hand, no Coulombian repulsion between protons. 
The nuclear surface is changing freely provided the deformation is not
exceeding the limit of $\beta_w$. The spectrum obtained in this way is quite
different from the harmonic one, specific to the $U(5)$ symmetry. This can be
seen not only from the fact the the ratio $E_{4^+}/E_{2^+}$ deviates from 2,
but also from that the functions associated to the states $0^+$ and $2^+$ in
the $E(5)$ and $U(5)$ symmetry descriptions respectively, are quite different
from each other.
For illustration we present, in Fig. 1, the graphs of the wave functions for
the two states as given by the corresponding $E(5)$ and $U(5)$ irreducible
representations. On the other hand, differences in wave functions manifest
themselves in 
the description of the transition probabilities which, as a matter of fact,
define
the objectives of this Section.

Here we suppose that the potential energy in the beta variable is depending on angular momentum in the following way:

\begin{eqnarray}
u(\beta)=\left \{\matrix{\beta^2,& \rm{if}&\;\;0\leq \beta <\infty,&\;\;L\leq 2,\cr  
                         0,     & \rm{if}&\;\;0\leq\beta \leq \beta_w,&\;\;L\geq 4,\cr
                         \infty,& \rm{if}&\;\;\beta_w < \beta <\infty,&\;\; L\geq 4.}\right.
\label{redpot}
\end{eqnarray}

The states of interest and their energies have the following expressions:
\begin{eqnarray}
|L^+_{n\tau}M\rangle &=& \sqrt{\frac{2n!}{\Gamma(n+\tau+5/2)}}\beta^\tau
L^{\tau +3/2}_n(\beta^2)e^{-\beta^2/2}G^{LM}_{n\tau}(\gamma,\Omega), \nonumber\\
E_{n\tau} &=& \frac{\hbar^2}{2B}(2n+\tau+5/2),\;\;(n,\tau)=(0,0),\;(0,1),\;\;L=2\tau,\nonumber\\
|L^+_{\xi,\tau}M\rangle &=& C_{\xi,\tau}\beta^{-3/2}J_{\tau+3/2}
(\beta x_{\xi,\tau}/\beta_w)G^{LM}_{\xi-1,\tau}(\gamma,\Omega),\nonumber\\
E_{\xi,\tau} &=& \frac{\hbar^2}{2B}\frac{x_{\xi,\tau}^2}{\beta^2_w},\;\;
(\xi,\tau )=(1,2),\;(1,3),\;(2,0).
\label{states}
\end{eqnarray}
The factor functions depending on the beta variable are solutions of
Eq.(\ref{epsf}) with the reduced potential given by Eq.(\ref{redpot}).
The equation for $\gamma$ deformation and Eulerian angles ($\Omega$)
has the solution $G^{LM}_{n\tau}$. Analytical solutions for this function for
an arbitrary $(n\tau)$ have been found in Ref.\cite{Gheor} while the matrix
elements  for monomials in quadrupole coordinates between these states are
given in Ref.\cite{Rad6}. For low lying states, recursion formulae determining the  gamma dependence of the wave functions were presented in Ref.\cite{Bes}.
For the sake of completeness we listed the needed functions $G^{LM}_{n\tau}$ in Appendix A.  

To calculate the $B(E2)$ values corresponding to the
electromagnetic transitions which are relevant for pointing out the specific
features for the phase transition from $U(5)$ to $O(6)$ symmetry, we need  an
expression for the transition operator. In the present work,
this operator is taken as being proportional to the quadrupole collective
coordinate which, in terms of intrinsic coordinates and Eulerian angles, has the expression:
\begin{equation}
T_{2M}=t\beta \Gamma_{2M}(\gamma,\Omega),\;\;\Gamma_{2M}(\gamma,\Omega)=  \left(\cos {\gamma}D^2_{M0}(\Omega)+\frac{1}{\sqrt{2}}\sin{\gamma}
\left(D^2_{M2}(\Omega)+D^2_{M,-2}(\Omega)\right)\right).
\label{transop}
\end{equation}
Using the convention of Rose \cite{Ros} for the Wigner Eckart theorem, the reduced transition probability between the states of angular momenta $L$ and 
$L'$ is obtained by  squaring the corresponding reduced matrix elements of the transition operator:
\begin{equation}
B(E2;L\to L')=
\left[\langle L^+_{\eta,\tau}||T_2||L^{'+}_{n,\tau}\rangle\right]^2,\;\;
\eta=0,1,2;\;\;n=0,1\;;\;\;\tau=1,2,3.
\label{BE2}
\end{equation}
The states involved in the above equation are those defined in 
Eq.(\ref{states}). 
There are some recent calculations \cite{Arias}, which use an anharmonic
structure for the transition operator. Also in the quoted paper, all states are 
multiplied by  a 'condensate' factor (see Eq.(14) of the quoted reference) which
contains an additional dependence on beta and gamma variables.
The role of this condensate factor consists of transforming the boson
transition operator with an anharmonic structure into a classical function
depending nonlinearly on $\beta$.

The reduced matrix elements of the transition operator can be written as a
product of two factors one due to the integration on the beta variable and one
reduced matrix element of the factor depending on $\gamma$ and $\Omega$
variables, involved in the transition  operator. The latter one can be
analytically evaluated.  For the cases considered here, one obtains:.\
\begin{eqnarray}                                         
\langle L^+_{\eta \tau}||T_2||L^{'+}_{n \tau '}\rangle &=&{\cal F}^{(\beta)}_{(\eta \tau;n\tau ')}
\langle G^L_{\eta \tau}||\Gamma_{2}||G^L_{n \tau '}\rangle \;\;,
\nonumber\\
\langle G^4_{12}||\Gamma_{2}||G^2_{11}\rangle  
&=&\sqrt{\frac{2}{7}},\;\;
\langle G^0_{20}||\Gamma_{2}||G^2_{11}\rangle =1,\nonumber\\
\langle G^0_{13}||\Gamma_{2}||G^2_{12}\rangle &=&\frac{1}{\sqrt{3}},\; \; 
\langle G^2_{01}||\Gamma_{2}||G^0_{00}\rangle =\frac{1}{\sqrt{5}}.
\label{redmat}
\end{eqnarray}
The formalism described so far was applied to $^{134}$Ba and $^{104}$Ru.
We present first the predictions for $^{134}$Ba.
Thus, results  for 4 excitation energy ratios and three
$B(E2)$ ratios are presented in the first column of Table I,
for $\beta_w=3.1$. How did we arrive at this value for the size of the square well potential?
In order to make a suitable choice we analyzed first the structure of 
the first states corresponding to the infinite square well potential. The best way to do that would be to diagonalize the model Hamiltonian in the basis
$\{|N\tau\alpha JM\rangle\}$ with the standard notation for the quantum numbers: the number of quadrupole bosons ($N$),the  seniority ($\tau$),the missing quantum number ($\alpha $), angular momentum ($J$) and its projection on the axis z 
($M$). That can be achieved, indeed, following the procedure we described in 
Ref.\cite{Gheor} for
the triaxial rotor Hamiltonian. Here we adopt another method, namely we 
calculate the overlap of the above quoted states with the lowest four 
vibrational states
from the basis $|N\tau\alpha J M\rangle$. From Figs. 2 and 3 one notices that for the chosen value $\beta_w=3.1$, the functions for the lowest $0^+$ and $2^+$ states in the two pictures, $E(5)$ and $U(5)$, are maximally overlapping, respectively.
Moreover, Fig.1 shows that for this value the vibrational states wave functions are very small although their zeros is
rigorously reached only at infinity.

\begin{table}[h]
\begin{tabular}{|c|c|c|c|c|c|c|}
\hline
 &present$^{a)}$&present${^b)}$&$^{134}$Ba (Exp.)&$\hskip0.5cm$E(5)$\hskip0.5cm$&Sextic model$^{c)}$&$\hskip0.5cm \beta^4\;\;^{d)}\hskip0.5cm$\\
\hline
$\frac{E_{4^+_{1,2}}-E_{0^+_{1,0}}}{E_{2^+_{1,1}}-E_{0^+_{1,0}}}$
&2.58  & 2.39  & 2.32 &  2.20 & 2.39&2.09\\
$\frac{E_{0^+_{2,0}}-E_{0^+_{1,0}}}{E_{2^+_{1,1}}-E_{0^+_{1,0}}}$
&3.71 & 3.47 & 3.57 & 3.03 & 3.68 &2.39\\
$\frac{E_{6^+_{1,3}}-E_{0^+_{1,0}}}{E_{2^+_{1,1}}-E_{0^+_{1,0}}}$
&4.46 & 4.2 & 3.66 & 3.59 & 3.70 &3.27\\
$\frac{E_{0^+_{2,0}}-E_{0^+_{1,0}}}{E_{0^+_{1,3}}-E_{0^+_{1,0}}}$
&0.83  & 0.83  & 0.97  & 0.84  & 0.99  & 0.73\\
$\frac{B(E2;4^+_{1,2}\to 2^+_{1,1})}{B(E2;2^+_{1,1}\to 0^+_{1,0})}$
&1.88 & 1.20 & 1.56(18)&1.68 & 1.70 & 1.82 \\
$\frac{B(E2;0^+_{2,0}\to 2^+_{1,1})}{B(E2;2^+_{1,1}\to 0^+_{1,0})}$
&0.28 & 0.55 & 0.42(12)& 0.86 & 1.03 & 1.41 \\
$\frac{B(E2;0^+_{1,3}\to 2^+_{1,2})}{B(E2;2^+_{1,1}\to 0^+_{1,0})}$
&3.11  & 2.76  &   & 2.21 & 2.12 & 2.52 \\
\hline
\end{tabular}
\caption{Results for the relevant excitation energies given in unites of
the excitation energy of the first $2^+$ state are listed. Also, the BE2 values characterizing the transitions $4^+_1\to 2^+_1,\;0^+_2\to 2^+_1$ and 
$0^+_3\to 2^+_2$ are given in units of $B(E2;2^+_1\to 0^+_1)$. 
$^{a)}$Data from the first column are obtained by assuming that the states $0^+_1$ 
and $2^+_1$ are pure 
vibrational states and taking for the states described by
Bessel functions of half integer order $\beta_w=3.1$ .$^{b)}$ In the second column the
results correspond to a different structure for the first states.
They are
obtained by diagonalizing the Hamiltonian (\ref{Hanha}) with x=0.2.
The other states involved in our calculations, correspond to $\beta_w=3.186.$
Experimental data, taken from Ref.\cite{Zam}, are collected in the third column. In the fourth column are the results corresponding to the $E(5)$ symmetry while in $^{c)}$ the fifth and $^{d)}$ sixth
columns, the results of sextic model from Refs.\cite{Lev}  and with $\beta^4$ potential \cite{Aria1} respectively, are presented for comparison.}
\end{table}

\begin{table}[h]
\begin{tabular}{|c|c|c|c|}
\hline
 &present&$^{104}$Ru (Exp.)&$\hskip0.5cm$E(5)$\hskip0.5cm$\\
\hline
$\frac{E_{4^+_{1,2}}-E_{0^+_{1,0}}}{E_{2^+_{1,1}}-E_{0^+_{1,0}}}$
&2.40  & 2.48  & 2.20 \\
$\frac{E_{6^+_{1,3}}-E_{0^+_{1,0}}}{E_{2^+_{1,1}}-E_{0^+_{1,0}}}$
&4.22 & 4.35 & 3.59 \\
$\frac{E_{8^+_{1,4}}-E_{0^+_{1,0}}}{E_{2^+_{1,1}}-E_{0^+_{1,0}}}$
&6.29  & 6.48  & 5.17      \\
$\frac{E_{10^+_{1,5}}-E_{0^+_{1,0}}}{E_{2^+_{1,1}}-E_{0^+_{1,0}}}$
&8.60 & 8.69 & 6.93             \\
$\frac{E_{0^+_{2,0}}-E_{0^+_{1,0}}}{E_{2^+_{1,1}}-E_{0^+_{1,0}}}$
&3.49 & 2.76 & 3.02 \\
$\frac{E_{0^+_{2,0}}-E_{0^+_{1,0}}}{E_{0^+_{1,3}}-E_{0^+_{1,0}}}$
&0.83  & 0.528  & 0.84  \\
$\frac{B(E2;4^+_{1,2}\to 2^+_{1,1})}{B(E2;2^+_{1,1}\to 0^+_{1,0})}$
&1.18 & 1.18(15) & 1.68 \\
$\frac{B(E2;0^+_{2,0}\to 2^+_{1,1})}{B(E2;2^+_{1,1}\to 0^+_{1,0})}$
&0.47 & 0.42(7) & 0.86 \\
$\frac{B(E2;0^+_{1,3}\to 2^+_{1,2})}{B(E2;2^+_{1,1}\to 0^+_{1,0})}$
&2.85  &   &  2.21 \\
\hline
\end{tabular}
\caption{Results for the relevant excitation energies given in unites of
the excitation energy of the first $2^+$ state are listed. Also, the BE2 values characterizing the transitions $4^+_1\to 2^+_1,\;0^+_2\to 2^+_1$ and 
$0^+_3\to 2^+_2$ are given in units of $B(E2;2^+_1\to 0^+_1)$. 
 Data from the first column are obtained by assuming that the states $0^+_1$
and $2^+_1$ are eigenstates of the Hamiltonian (\ref{Hanha}) with x=0.15, while the remaining ones are
described by the Bessel functions of half integer order. For all states the order parameter
has the value $\beta_w=3.17$ .
 In the second column the experimental data taken from Ref.\cite{Gia,Bla} are given.
The results from the third column correspond to the E(5) symmetry. }
\end{table}

Having in mind the $E(5)$ wave functions structure revealed by Figs.2 and 3
one may think of improving the agreement with the experimental data by using 
for the vibrational states some anharmonic components.
Thus, the reduced energies $\epsilon$ for the first states $0^+$ and $2^+$
are obtained by diagonalizing the anharmonic Hamiltonian:
\begin{equation}
H_b=({\hat N} +\frac{5}{2})+x\left((b^{\dagger}_2b^{\dagger}_2)_0+(b_2b_2)_0\right),
\label{Hanha}
\end{equation}
where ${\hat N}$ denotes the number of quadrupole bosons which hereafter will be denoted by  $b^{\dagger}_{2\mu}$ and $b_{2\mu}$.
Since our application has only an illustrative value we restrict the dimension of the diagonalization space to 2.
The final results for energies and wave functions are:
\begin{eqnarray}
\epsilon _{0^{+}} &=&\frac{7}{2}-u,\quad \epsilon _{2^{+}}=\frac{9}{2}-v, 
\nonumber \\
u &=&\sqrt{1+2x^{2}},\quad v=\sqrt{1+\frac{14}{5}x^{2}}, \\
|0^+\rangle   &=&X_0|0\rangle +Y_0 \frac{1}{\sqrt{2}}(b^{\dagger}_2b^{\dagger}_2)_0|0\rangle\;,\nonumber\\
|2^+\mu \rangle   &=&X_2b^{\dagger}_{2\mu}|0\rangle +Y_2 \sqrt{\frac{5}{14}}
(b^{\dagger}_2b^{\dagger}_2)_0b^{\dagger}_{2\mu}|0\rangle\;.
\end{eqnarray}
The amplitudes of the states yielded by diagonalization have the expressions:

\begin{eqnarray}
X_{0} &=&\frac{x}{\sqrt{u^{2}-u}}\;,\quad Y_{0}=\sqrt{\frac{u-1}{2u}}\;, 
\nonumber \\
X_{2} &=&\sqrt{\frac{7}{5}}\frac{x}{\sqrt{v^{2}-v}},\;\quad Y_{2}=\sqrt{%
\frac{v-1}{2v}}.
\end{eqnarray}

How do the multi-boson states components, involved in the states $0^+$ and $2^+$, affect the results for energies and transition probabilities, can be seen in the second column of Table I. The maximal overlap of
 the  anharmonic states $0^+$ and $2^+$ with the corresponding $E(5)$ states requires changing $\beta_w$ to the value of 3.186.
Comparing the results obtained in this way with the corresponding 
experimental data one may say that  anharmonicities improve the agreement with the experimental data.
Comparing the present results with the ones obtained through different 
formalisms, one remarks that our procedure is the only one succeeding to provide a reasonable agreement with the
experimental data for both excitation energy and the $E2$ 
decay probability of the state $0^+_{2,0}$
to the first $2^+$ state. In this context it is noteworthy the fact that
describing the head states of the excited bands is a decisive test for any
nuclear structure model.

Another nucleus  whose behavior reclaims an E(5) symmetry is $^{104}$Ru.
The even isotopes of Ru have been experimentally studied in Ref.\cite{Gia}.
Their global properties which are relevant for the context of E(5) symmetry have
been analyzed within the framework of IBA model in Ref.\cite{Fra}. Thus, the A dependence
of the excitation energy for $0^+_2$ suggests a transitional behavior near
$^{104}$Ru,
although this is not reflected in the two-neutron separation energies.
Moreover, averaging the IBA Hamiltonian on a coherent state, the resulting
function of deformation and boson number admits N=8,
which corresponds to $^{104}$Ru,
as a critical value.
Indeed,  the energy function has a minimum in $\beta=0$ for $N<8$,
while for $N\geq8$ the minimum value is reached for a nonvanishing $\beta$.
The IBA description uses  a 5 parameter boson Hamiltonian and an anharmonic
structure for the
transition operator. Four of the five parameters are fixed so that an overall
agreement for all $^{A}_{44}$Ru isotopes is obtained.  Predictions for energies and
transition probabilities are in very good agreement with the corresponding
experimental data.

Results of our calculations for $^{104}$Ru are collected in Table II.
Here we considered
for the first two states, $0^+$ and $2^+$, an anharmonic structure by diagonalizing
the boson Hamiltonian (\ref{Hanha}). The higher excited states are described by the
E(5) formalism.
As in the previous case there are two parameters involved, $\beta_w$ and $x$.
The first one is fixed so that the overlap of vibrational states and E(5) states have
a maximal overlap. This way of fixing $\beta_w$ is justified by the fact that at
critical point
the specific features of the extreme symmetries, $SU(5)$ and $SO(6)$,
are maximally mixed.
It is worth noting that by contrast to the E(5) formalism where the normalized
energies do not depend on $\beta_w$, here the results for both transition probabilities
and energies are depending on this parameter.
The other parameter ($x$) was fixed so that the most sensitive transition
probability is reproduced. For comparison in Table II we give also the results obtained with
the E(5) formalism. Note that excepting the energy of $0^+_2$ all other data are fairly well reproduced.
Moreover, from Table II it is conspicuous that the present results are much closer to the experimental data
than those provided by the E(5) description.

Concluding, the features mentioned above support our hypothesis asserting that 
the critical value for the Hamiltonian order parameters (which is $\beta_w$)
is angular momentum dependent.
\section{Relationship of spherical vibrator and E(5) states.}
\label{sec:level4}
As we stated already, the infinite 5 dimensional well is simulating the situation met for some transitional nuclei of a flat potential in the beta variable. 
This is, indeed, a very particular picture taking into account that the $O(6)$ nuclei are soft only in the $\gamma$ variable. If that is the case, then one suspects that there are some mathematical features which vindicate this approximation. 

Also we note that the procedure used for obtaining the quantized energy for $U(5)$ and $E(5)$ symmetries are somehow different. While for $U(5)$, one requires that the wave function has a regular behavior at infinity, for the $E(5)$ states the condition is to
vanish for the wall coordinate. 

Energies provided by the zeroes of the Bessel  functions of half integer order are organized in bands labeled by the
quantum number $\xi$, which numbers the above cited zeros.
For example, the ground band comprises states with $L=2\tau$, and $\xi=1$.  On the other hand the spherical vibrator states are classified in bands according to a different rule, namely the ground band is the set of $n=0$ states of highest seniority, the beta band is the second highest seniority band, etc. According to these schemes of classification one may put in correspondence
the 
states of spherical vibrator ground band with those of the ground band in the E(5) picture. Thus, to the highest seniority $\xi=1$ states would correspond the highest seniority $n=0$ states. Such a correspondence is also supported by Fig. 1 showing that the corresponding states have similar allure, although they differ in magnitude.

Concerning the ground band energies  for the spherical vibrator another remark is worth to be mentioned.
Since the polynomial $L^{\tau+3/2}_1(x)$ is linear in $x$, it has only one zero at
\begin{equation}
x=\tau +5/2
\end{equation}
which can describe the ground band energies by considering
\begin{equation}
\epsilon_{0\tau}=\beta^2_0.
\label{epsibe}
\end{equation}
with $\beta_0$ denoting the zero given by Eq.(4.1) where $x$ is replaced by 
$\beta_0^2$.
Inspecting Eq.(\ref{Sch}), one sees that for these wave function zeros, the second order derivative is also vanishing. Reversely, the solution
of the equation (\ref{epsibe}) in $\beta_0$ is a zero for the wave function. This reflects a classical property of this potential consisting in that the solutions for
 Eq.(\ref{epsibe}) provides the turning points of the classical trajectory. 
The comments from above raise the question whether it is possible that all energy levels from the spherical vibrator picture can be given as zeros of some generalized Laguerre polynomials of a
contracted variable. Moreover, is it  possible
to describe the vibrator states in terms of Bessel functions of half integer
order? The answers of these questions are given by using the asymptotic behavior of generalized Laguerre polynomials \cite{Bate} which approximate quite well
the Bessel functions according to the formula.  

\begin{eqnarray}
\widetilde{L}(s,n,z)&\equiv&\frac{n!}{\Gamma(n+s+1)}
\left(\frac{z}{2}\right)^s\exp{\left(-\frac{z^2}{4+8n+4s}\right)}
L^s_n\left(\frac{z^2}{2+4n+2s}\right),
\nonumber\\
\widetilde{L}(s,n,z)&\approx &J_s(z).
\label{approxJ}
\end{eqnarray}

The function $\widetilde{L}$ approximates exceptionally well the Bessel function of half integer for large values of $n$. The approximation is, however, reasonable good even for small values of $n$.
 For Illustration we present in Table III, the values of the deviations from unity of the ratio of $\widetilde{L}(s,n,z)$ and $J_s(z)$ for few values of $n$.
\begin{equation}
R(s,n,z)=\left[\widetilde{L}(s,n,z)/J_s(z)-1\right].
\label{ratioR}
\end{equation}
Actually this equation fixes the conditions under which the E(5) energies and
functions might be described in terms of U(5) energies and states, respectively.
  
\begin{table}[h]
\begin{tabular}{|c|c|c|c|c|c|c|c|c|c|c|c|}
\hline
n& \multicolumn{11}{c|}{$\tau+3/2$}\\
\hline
 &3/2& 5/2 & 7/2 & 9/2& 11/2 & 13/2 & 15/2 & 17/2 &
 19/2 & 21/2 & 23/2 \\
\hline
0 & 26.4&9.8&4.7&2.6&1.6&1.1&0.7&0.5&0.4&0.3&0.2\\
1 & 8.1 & 3.9 &2.2 &1.4 & 0.9& 0.7& 0.5&0.4&0.3&0.2&0.2\\
2&3.9&2.1&1.3&0.9&0.6&0.5&0.3&0.3&0.2&0.2&0.1\\                    
3& 2.2& 1.3 & 0.9 & 0.6 & 0.4 & 0.3 & 0.3 & 0.2 & 0.2 & 0.1 & 0.1\\
4 & 1.5 & 0.9 & 0.6 & 0.4 & 0.3 & 0.2 & 0.2 & 0.2 & 0.1 & 0.1 & 0.09 \\
5 & 1.0 & 0.7 & 0.5 & 0.3 & 0.3 & 0.2 & 0.2 & 0.1 & 0.1 & 0.09& 0.07\\
\hline
\end{tabular}
\caption{The function $R(s,n,z)$ given by Eq.(\ref{ratioR}), 
 multiplied by a factor of $10^3$, was calculated for $s=\tau+3/2,\;\;z=2$ and
several values of the quantum number $n$ .}
\end{table}

Let us replace $z=\sqrt{\epsilon}\beta$ in Eq. (\ref{approxJ}) and denote
by $y_{\xi,n,\tau}$ the $\xi$-th zero of the generalized Laguerre polynomial
$L^{\tau+3/2}_n(y)$. Then the corresponding energy is:
\begin{equation}
E_{\xi,n,\tau}=\frac{\hbar^2}{2B}(4n+2\tau+5)\frac{y_{\xi,n,\tau}}{\beta^2_w}.
\end{equation}
and the normalized excitation energies  
\begin{equation}
R_{n,\tau}=\frac{E_{1,n,\tau}-E_{1,n,0}}{E_{1,n,1}-E_{1,n,0}}.
\label{rationtau}
\end{equation}
are readily calculated.
For large values of $n$ this ratio must converge to the ratio determined by the energies given by the zeros of the Bessel function of half integer order.
\begin{equation}
R_{\tau}=\frac{x_{1,\tau}^2-x_{10}^2}{x_{11}^2-x_{10}^2}.
\label{ratiotau}
\end{equation}
In Table IV we list the values of ratio (\ref{rationtau}) and (\ref{ratiotau}).
From there, one may see that although the convergence is reached for large
values of $n$, even the low values of $n$ produce energy ratios which are close to the E(5) limit.

\begin{table}[h]
\begin{tabular}{|c|c|c|c|c|c|c|c|}
\hline
  &\multicolumn{7}{c|}{$\tau$}\\
\hline
 &  0   &   1   &   2   &   3   &   4   &   5   &   6     \\
\hline
$J_{\tau +3/2}$
 &  0   &   1   & 2.19859 & 3.58982 & 5.16941 & 6.93412 & 8.8814   \\
$L^{\tau +3/2}_1$
 &  0   &   1   & 2.25    & 3.75    &  5.5    &   7.5   &  9.75     \\
$L^{\tau +3/2}_2$
 &  0   &   1   & 2.23793 & 3.71562 & 5.43480 & 7.39708 & 9.60389    \\
$L^{\tau +3/2}_3$
 &  0   &   1   & 2.22819 & 3.68612 & 5.37567 & 7.29876 & 9.45723     \\
$L^{\tau +3/2}_4$
 &  0   &   1   & 2.22136 & 3.66488 & 5.33204 & 7.22458 &9.34431   \\
$L^{\tau +3/2}_5$
 &  0   &   1   & 2.21657 & 3.64966 & 5.30023 & 7.16969 &9.25963       \\
$L^{\tau +3/2}_6$
 &  0   &   1   & 2.21310 & 3.63852 & 5.27666 & 7.128 & 9.19550     \\
$L^{\tau +3/2}_{100}$
 &  0   &   1   & 2.19872 & 5.59027 & 5.17046 & 6.93615 & 8.88490\\
\hline
\end{tabular}
\caption{The ratios $R_{n,\tau}$ given by Eq.(\ref{rationtau}) are calculated 
for several values on $(n,\tau)$. In the first column are given the generalized
Laguerre polynomials whose first zero determine the involved energies.
In the first row the $E(5)$ limit, i.e. the ratio $R_{\tau}$ given by Eq. 
(\ref{ratiotau}), is listed. The corresponding energies are zeros for the Bessel function given in the first column.}
\end{table}

Concluding, although the generalized Laguerre polynomials and Bessel functions of half integer order account for different symmetries, according to Eq. (\ref{approxJ}) they approximate each other for large values of $n$. However, in virtue of Eq.(2.11) the generalized Laguerre polynomials are eigenstate of one of the $SU(1,1)$ generators. On the other hand 
this group may be contracted to an Euclidean one. 
Therefore, the asymptotic behavior reflected in Eq.(\ref{approxJ}) is supported by the group relationship mentioned above.
On the other hand the fact that asymptotically the generalized Laguerre
polynomials can be approximated by Bessel functions of half integer order is a nice justification of replacing the potential in $\beta$ characterizing the transitional U(5)-O(6) nuclei, by a five dimensional infinite well potential.
  
\section{Summary and Conclusions}
\label{sec:level5}
The main results obtained in the present paper can be summarized as follows.
We suppose that in the critical point of the $U(5)-O(6)$ transition the
potential in $\beta$ is angular momentum dependent and given by Eq.(\ref{redpot}). 
To be more concrete, the ground and the first $2^+$ states are described by a 
spherical vibrator Hamiltonian while the higher states by a Hamiltonian involving a five dimensional infinite well potential in $\beta$. This simulates a scenario where
the higher states undergo the phase transition while the first states
still behave as spherical vibrator states. Thus,
the assigned wave functions are those from Eq.(\ref{states}).
At the next stage we considered some anharmonicities in the structure of the
spherical states. The results obtained for energies and transition
probabilities are collected in Table I for $^{134}$Ba and in Table II for $^{104}$Ru.

From Table I one notices a good description for the state $0^+_{2,0}$. This
supports our hypothesis for a gradual transition from a spherical to
a
$E(5)$ symmetry. The predictions shown in Table II for $^{104}$Ru suggest a very good
agreement with the experimental data. Moreover, results of our calculations
are much closer to the experimental data than
those obtained with the E(5) symmetry formalism.
The present results encourage us for searching for new nuclei whose  behavior
is compatible with an E(5) symmetry in some states but with U(5) and possible
with O(6) symmetries in the remaining states.

The connection between the basis states specific to the two symmetries was
discussed in Section 4. The relation  (\ref{approxJ}), which holds in the
asymptotic region for the quantum number $n$ may be considered as a
mathematical frame for the phase transition and moreover suggests a
possible way of relating the descriptions of the two phases.

\section{ Appendix A}
\label{sec:level6}
Here we list the normalized functions $G_{n\tau}$ which we need in Section 3.
\begin{eqnarray}
G^{00}_{00}(\gamma,\Omega )&=&G^{00}_{10}(\gamma,\Omega )=\frac{1}{4\pi},\;\;
G^{00}_{03}(\gamma,\Omega )=\frac{1}{4\pi}\sqrt{\frac{3}{2}}\cos {3\gamma},\nonumber\\
G^{2M}_{01}(\gamma,\Omega )&=&\frac{1}{4\pi}\sqrt{\frac{5}{2}}
\left[\cos{\gamma}D^2_{M0}(\Omega)
+\frac{\sin {\gamma}}{\sqrt{2}}\left(D^2_{M2}(\Omega)+D^2_{M,-2}(\Omega)\right) \right],\nonumber\\
G^{2M}_{02}(\gamma,\Omega )&=&\frac{1}{4\pi}\sqrt{\frac{5}{2}}\left[
\cos{2\gamma}D^2_{M0}(\Omega)-
\frac{\sin {2\gamma}}{\sqrt{2}}\left(D^2_{M2}(\Omega)+D^2_{M,-2}(\Omega)\right) \right],\nonumber\\
G^{4M}_{02}(\gamma,\Omega )&=&\frac{1}{8\pi\sqrt{2}}\left[\left(6\cos^2\gamma +\sin^2\gamma \right)D^4_{M0}(\Omega)+\frac{15}{\sqrt{2}}\sin {2\gamma}
\left(D^4_{M2}(\Omega)+D^4_{M,-2}(\Omega)\right)\right.\nonumber\\
&&\left.+\frac{35}{\sqrt{2}}\sin ^2\gamma 
\left(D^4_{M4}(\Omega)+D^4_{M,-4}(\Omega)\right) \right].\nonumber\\
\end{eqnarray}

\begin{figure}[tb]
\par
\begin{center}
\leavevmode
\epsfxsize = 16 cm \epsffile[0 120 590 816]{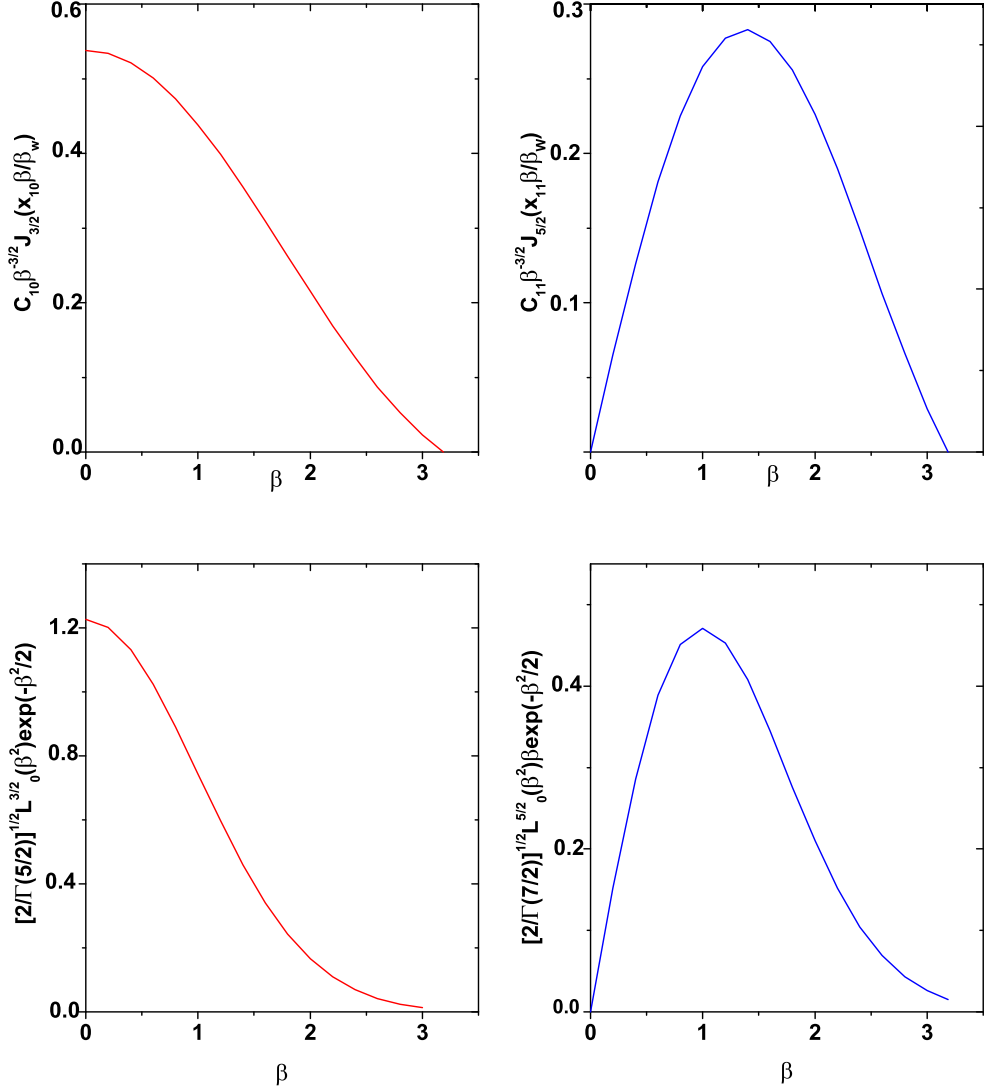}
\end{center}
\caption{
(Color online) First row:the wave functions describing the first 
two states, $0^+$-left panel and $2^+$-right panel, in a 5 dimensional infinite well. Second row:the wave functions describing the first two states, $0^+$-left panel and $2^+$-right panel, in a 5 dimensional oscillator well.
}
\label{Fig. 3}
\end{figure}

\clearpage

\begin{figure}[tb]
\par
\begin{center}
\leavevmode
\epsfxsize = 16 cm \epsffile[0 120 590 816]{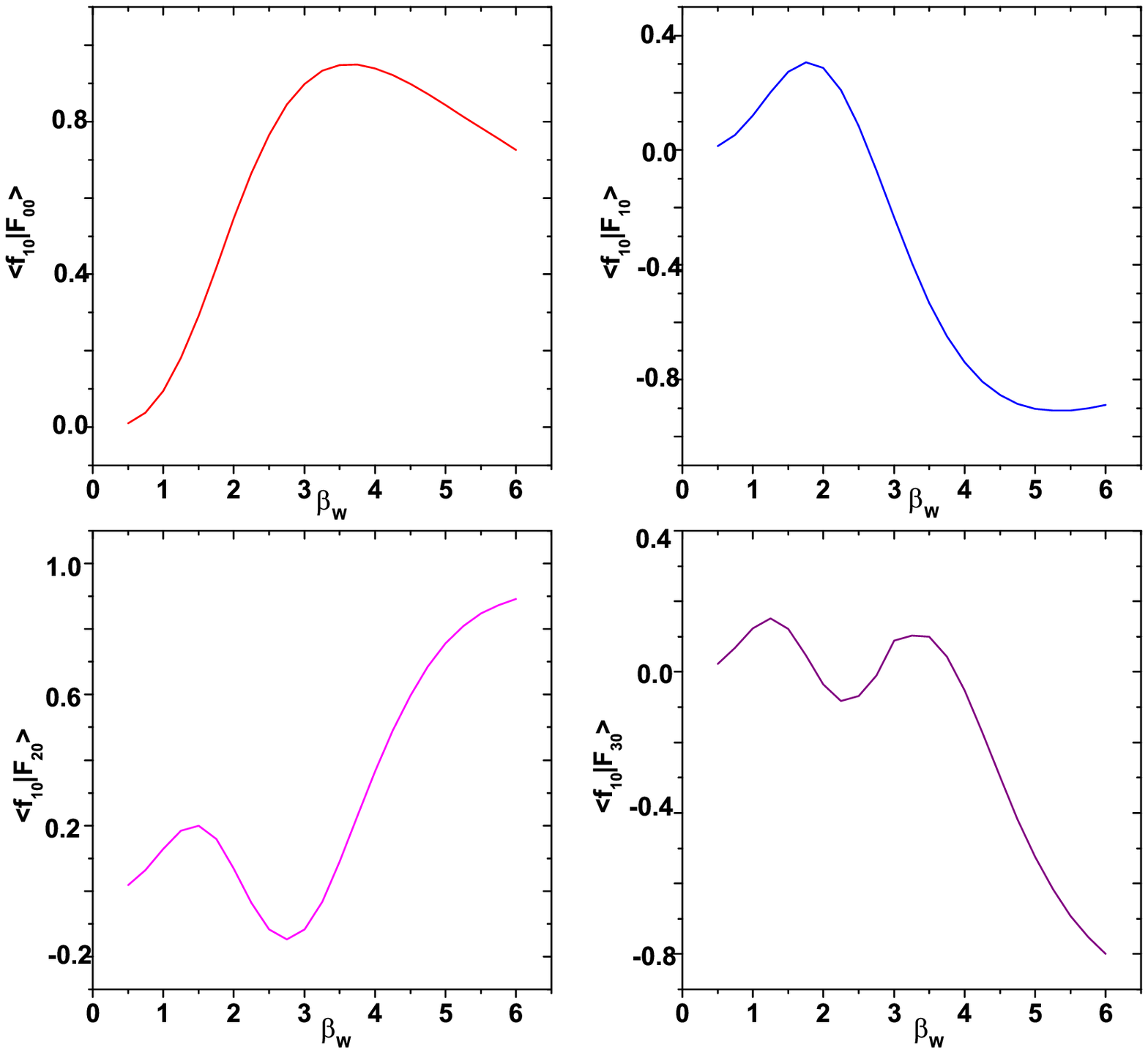}
\end{center}
\caption{
(Color online)The overlaps of the functions
$f_{\xi,\tau}$ with $\xi=1,\tau=0$ given by Eq.(2.20) and $F_{n\tau}$ (Eq. (2.9)) with ($n,\tau$)=(0,0)(upper-left panel),(1,0)(upper-right panel),(2,0)
(bottom-left panel),(3,0)(bottom-right panel).
}
\label{Fig. 2}
\end{figure}
\clearpage

\begin{figure}[tb]
\par
\begin{center}
\leavevmode
\epsfxsize = 16 cm \epsffile[0 120 590 816]{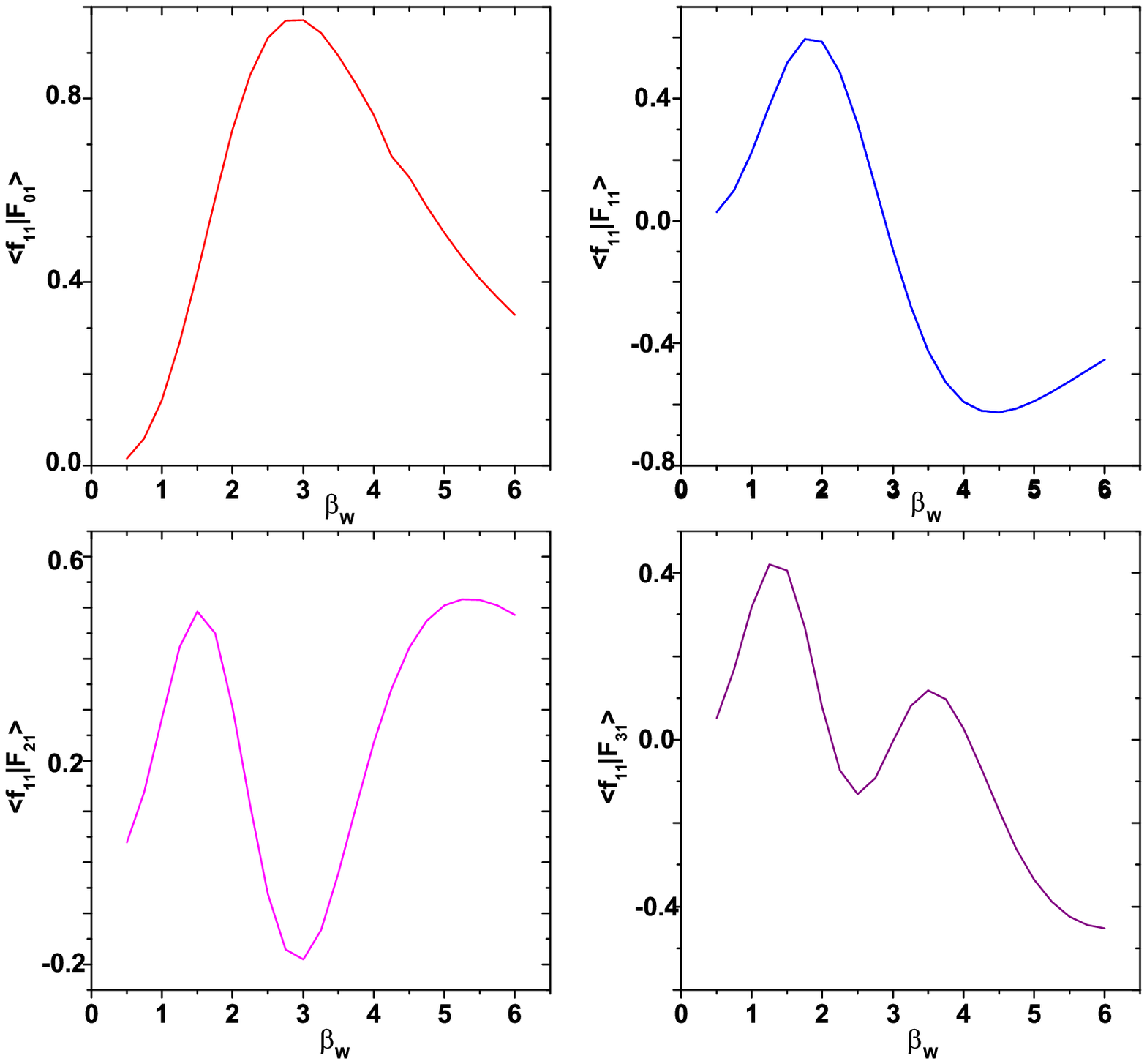}
\end{center}
\caption{
(Color online) The overlaps of the functions
$f_{\xi,\tau}$ with $\xi=1,\tau=1$ given by Eq. (2.20) and $F_{n\tau}$ (2.9) with ($n,\tau$)=(0,1)(upper-left panel),(1,1)(upper-right panel),(2,1)(bottom-left panel),(3,1)(bottom-right panel)
}
\label{Fig. 3}
\end{figure}

\clearpage

\end{document}